\documentclass[twocolumn, showpacs, amsmath,amssymb]{revtex4} 
\usepackage{epsfig, amssymb}
\begin{document}
\preprint{xxx}

\title{Graphical description of the action of local Clifford transformations on graph states}

\author{Maarten Van den Nest}
\email{maarten.vandennest@esat.kuleuven.ac.be} \author{Jeroen Dehaene}\author{Bart De Moor}
 \affiliation{Katholieke Universiteit Leuven, ESAT-SCD, Belgium.}
\date{\today}

\begin{abstract}
We translate the action of local Clifford operations on graph states into transformations on their
associated graphs - i.e. we provide transformation rules, stated in purely graph theoretical terms,
which completely characterize the evolution of graph states under local Clifford operations. As we
will show, there is essentially one basic rule, successive application of which generates the orbit
of any graph state under local unitary operations within the Clifford group.
\end{abstract}

\maketitle


\section{Introduction}

Stabilizer states and, more particularly, graph states, and (local) unitary operations in the
Clifford group have been studied extensively and play an important role in numerous applications in
quantum information theory and quantum computing. A stabilizer state is a multiqubit pure state
which is the unique simultaneous eigenvector of a complete set of commuting observables in the
Pauli group, the latter consisting of all tensor products of Pauli matrices and the identity (with
an additional phase factor). Graph states are special cases of stabilizer states, for which the
defining set of commuting Pauli operators can be constructed on the basis of a mathematical graph.
The Clifford group consists of all unitary operators which map the Pauli group to itself under
conjugation. As the closed framework of stabilizer theory plus the Clifford group turns out to have
a relatively simple mathematical description while maintaining a sufficiently rich structure, it
has been employed in various fields of quantum information theory and quantum computing: in the
theory of quantum error-correcting codes, the stabilizer formalism is used to construct so-called
stabilizer codes which protect quantum systems from decoherence effects \cite{Gott}; graph states
have been used in multipartite purification schemes \cite{graphbriegel} and a measurement-based
computational model has been designed which uses a particular graph state, namely the cluster
state, as a universal resource for quantum computation - the one-way quantum computer
\cite{1wayQC}; (a quotient group of) the Clifford group has been used to construct performant
mixed-state entanglement distillation protocols \cite{loc_per_ent_dist}; most recently, graph
states were considered in the context of multiparticle entanglement: in \cite{entgraphstate} the
entanglement in graph states was quantified and characterized in terms of the Schmidt measure.


The goal of this paper is to translate the action of local Clifford operations on graph states into
transformations on their associated graphs - that is, to derive transformation rules, stated in
purely graph theoretical terms, which completely characterize the evolution of graph states under
local Clifford operations. The main reason for this research is to provide a tool for studying the
local unitary (LU) equivalence classes of stabilizer states or, equivalently, of graph states
\footnote{We will show in section III that each stabilizer state is equivalent to a graph state
under the local Clifford group.} - since the quantification of multi-partite pure-state
entanglement is far from being understood and a treatise of the subject in its whole is extremely
complex, it is appropriate to restrict oneself to a more easily manageable yet nevertheless
interesting subclass of physical states, as are the stabilizer states. The ultimate goal of this
research is to characterize the LU-equivalence classes of stabilizer states, by finding suitable
representatives within each equivalence class and/or constructing a complete and minimal set of
local invariants which separate the stabilizer state orbits under the action of local unitaries. We
believe that the result in this paper is a first significant step in this direction.

In section IV, we will show that the orbit of any graph state under local unitary operations within
the Clifford group is generated by repeated application of essentially one basic graph
transformation rule. The main tool for proving this result will be the representation of the
stabilizer formalism and the (local) Clifford group in terms of linear algebra over $GF(2)$, where
$n$-qubit stabilizer states are represented as $n$-dimensional linear subspaces of ${\Bbb
Z}_2^{2n}$ which are self-orthogonal with respect to a symplectic inner product \cite{Cal_Orth_Geo,
Gott} and where Clifford operations are the symplectic transformations of ${\Bbb Z}_2^{2n}$
\cite{loc_per_ent_dist, stab_clif_GF2}.

This paper is organized as follows: in section II, we start by recalling the notions of stabilizer
states, graph states and the (local) Clifford group and the translation of these concepts into the
binary framework. In section III, we then show (constructively) that each stabilizer state is
equivalent to a graph state under local Clifford operations, thereby rederiving a result of
Schlingemann \cite{stabgraphcode}. Continuing within the class of graph states, in section IV we
introduce our elementary graph theoretical rules which correspond to local Clifford operations and
prove that these operations generate the orbit of any graph state under local Clifford operations.

\section{Preliminaries}

\subsection{Stabilizer states, graph states and the (local) Clifford group}

Let ${\cal G}_n$ denote the Pauli group on $n$ qubits, consisting of all $4\cdot 4^n$ $n$-fold
tensor products of the form $\alpha\ v_1 \otimes v_2\otimes \dots \otimes v_n$, where $\alpha \in
\{ \pm 1, \pm i\}$ is an overall phase factor and the $2 \times 2$-matrices $v_i$ $(i=1,\dots, n)$
are either the identity $\sigma_0$ or one of the Pauli matrices \[ \sigma_x= \left(
\begin{array}{cc}0 & 1\\1 & 0 \end{array}\right),\ \sigma_y = \left(
\begin{array}{cc}0 & -i\\i & 0 \end{array}\right),\ \sigma_z=\left(
\begin{array}{cc}1 & 0\\0 & -1 \end{array}\right).  \]
The Clifford group ${\cal C}_n$ is the normalizer of ${\cal G}_n$ in $U(2^n)$, i.e. it is the group
of unitary operators $U$ satisfying $U {\cal G}_n U^{\dagger} = {\cal G}_n$. We shall be concerned
with the local Clifford group ${\cal C}_n^l$, which is the subgroup of ${\cal C}_n$ consisting of
all $n$-fold tensor products of elements in ${\cal C}_1$.

An $n$-qubit stabilizer state $|\psi\rangle$ is defined as a simultaneous eigenvector with
eigenvalue 1 of $n$ commuting and independent \footnote{This means that no product of the form
$M_1^{x_1}\dots M_n^{x_n}$, where $x_i \in \{0,1\}$, yields the identity except when all $x_i$ are
equal to zero.} Pauli group elements $M_i$. The $n$ eigenvalue equations $M_i|\psi\rangle =
|\psi\rangle$ define the state $|\psi\rangle$ completely (up to an arbitrary phase). The set ${\cal
S} := \{M \in {\cal G}_n| M|\psi\rangle = |\psi\rangle\}$ is called the stabilizer of the state
$|\psi\rangle$. It is a group of $2^n$ commuting Pauli operators, all of which have a real overall
phase $\pm 1$ and the $n$ operators $M_i$ are called generators of ${\cal S}$, as each $M \in {\cal
S}$ can be written as $M = M_1^{x_1}\dots M_n^{x_n}$, for some $x_i \in \{0,1\}$. The so-called
graph states \cite{1wayQC, graphbriegel} constitute an important subclass of the stabilizer states.
A graph \cite{graphtheory} is a pair $G=(V, E)$ of sets, where $V$ is a finite subset of $\Bbb{N}$
and the elements of $E$ are 2-element subsets of $V$. The elements of $V$ are called the vertices
of the graph $G$ and the elements of $E$ are its edges. Usually, a graph is pictured by drawing a
(labelled) dot for each vertex and joining two dots $i$ and $j$ by a line if the correspondig pair
of vertices $\{i,j\} \in E$. For a graph with $|V|=n$ vertices, the adjacency matrix $\theta$ is
the symmetric binary $n\times n$-matrix where $\theta_{ij}=1$ if $\{i,j\} \in E$ and
$\theta_{ij}=0$ otherwise. Note that there is a one-to-one correspondence between a graph and its
adjacency matrix. Now, given an $n$-vertex graph $G$ with adjacency matrix $\theta$ one defines $n$
commuting Pauli operators
\[K_j = \sigma^{(j)}_x \prod_{k=1}^n \left(\sigma_z^{(k)}\right)^{\theta_{kj}},\] where
$\sigma^{(i)}_x, \sigma^{(i)}_y, \sigma^{(i)}_z$ are the Pauli operators which have resp.
$\sigma_x, \sigma_y, \sigma_z$ on the $i$th position in the tensor product and the identity
elsewhere. The graph state $|\psi_{\mu_1\mu_2\dots\mu_n}(G)\rangle$, where $\mu_i \in \{0,1\}$, is
then the stabilizer state defined by the equations \[(-1)^{\mu_j}K_j\
|\psi_{\mu_1\mu_2\dots\mu_n}(G)\rangle = |\psi_{\mu_1\mu_2\dots\mu_n}(G)\rangle.\] Since one can
easily show that the $2^n$ eigenstates $|\psi_{\mu_1\mu_2\dots\mu_n}(G)\rangle$ are equal up to
local unitaries in the Clifford group, it suffices for our purposes to choose one of them as a
representative of all graph states associated with $G$. Following the literature
\cite{entgraphstate}, we denote this representative by $|G\rangle := |\psi_{00\dots 0}(G)\rangle$.
Furthermore, if the adjacency matrices of two graphs $G$ and $G'$ differ only in their diagonal
elements, the states $|G\rangle $ and $|G'\rangle$ are equal up to a local Clifford operation,
which allows for an a-priori reduction of the set of graphs which needs to be considered in the
problem of local unitary equivalence. The most natural choice is to consider the class $\Theta
\subseteq \Bbb{Z}_2^{n\times n}$ of adjacency matrices which have zeros on the diagonal. These
correspond to so-called simple graphs, which have no edges of the form $\{i,i\}$ or, equivalently,
none of the points is connected to itself with a line. From this point on, we will only consider
graph states which are associated with simple graphs.

\subsection{The binary picture}

It is well-known \cite{Cal_Orth_Geo, Gott, QCQI} that the stabilizer formalism can be translated
into a binary framework, which essentially exploits the homomorphism between ${\cal G}_1, \cdot$
and $\Bbb{Z}_2^{2},+$ which maps $\sigma_0=\sigma_{00} \mapsto 00$, $\sigma_x=\sigma_{01} \mapsto
01$, $\sigma_z=\sigma_{10} \mapsto 10$ and $\sigma_y=\sigma_{11} \mapsto 11$. In $\Bbb{Z}_2^{2}$
addition is to be performed modulo 2. The generalization to $n$ qubits is defined by
$\sigma_{u_1v_1}\otimes\dots\otimes\sigma_{u_nv_n} = \sigma_{(u_1\dots u_n|v_1\dots v_n)} \mapsto
(u_1\dots u_n|v_1\dots v_n) \in \Bbb{Z}_2^{2n}$, where $u_i, v_i \in \{0, 1\}$. Thus, an $n$-fold
tensor product of Pauli matrices is represented as a $2n$-dimensional binary vector. Note that with
this encoding one loses the information about the overall phases of Pauli operators. For now, we
will altogether disregard these phases and we will come back to this issue later in this paper.

In the binary language, two Pauli operators $\sigma_a$ and $\sigma_b$, where $a, b \in
\Bbb{Z}_2^{2n}$, commute iff $a^T P b = 0$, where the $2n\times 2n$-matrix $P = \left[
\begin{array}{cc} 0 & I \\ I& 0 \end{array} \right]$ defines a symplectic inner product on
the space $\Bbb{Z}_2^{2n}$. The stabilizer of a stabilizer state then corresponds to an
$n$-dimensional linear subspace of $\Bbb{Z}_2^{2n}$ which is its own orthogonal complement with
respect to this symplectic inner product. Given a set of generators of the stabilizer, we assemble
their binary representations as the columns of a full rank $2n\times n$-matrix $S$, which satisfies
$S^TPS=0$ from the symplectic self-orthogonality property. The entire stabilizer subspace consists
of all linear combinations of the columns of $S$, i.e. of all elements $Sx$, where $x \in
\Bbb{Z}_2^{n}$. The matrix $S$, which is referred to as a generator matrix for the stabilizer, is
of course non-unique. A change of generators amounts to multiplying $S$ to the right with an
invertible $n\times n$-matrix, which performs a basis change in the binary subspace. Note that a
graph state which corresponds to a graph with adjacency matrix $\theta$, has
a generator matrix $S = \left [ \begin{array}{c} \theta\\
I \end{array}\right]$. Finally, it can be shown \cite{stab_clif_GF2, loc_per_ent_dist} that, as we
disregard overall phases, Clifford operations are just the symplectic transformations of
$\Bbb{Z}_2^{2n}$, which preserve the symplectic inner product, i.e. they are the $2n\times
2n$-matrices $Q$ which satisfy $Q^TPQ =P$. As local Clifford operations act on each
qubit separately, they have the additional block structure $Q = \left [ \begin{array}{cc} A&B\\
C&D
\end{array}\right]$, where the $n\times n$-blocks $A, B, C, D$ are diagonal. In this case, the
symplectic property of $Q$ is equivalent to stating that each submatrix $\left [
\begin{array}{cc} A_{ii} & B_{ii}
\\ C_{ii}& D_{ii}\end{array} \right ]$, which acts on the $i$th
 qubit, is invertible. The group of all such $Q$ will be denoted by $C^l$.

Thus, in the binary stabilizer framework, two stabilizer states $|\psi\rangle$ and $|\psi'\rangle$
with generator matrices $S$ and $S'$ are equivalent under the local Clifford group iff
\footnote{Here we have used the fact that $|\psi\rangle$ and $|\psi'\rangle$ are equivalent under
the local Clifford group iff they have equivalent stablizers ${\cal S}$, ${\cal S'}$, i.e. iff
there exists a local Clifford operation $U$ s.t. $U{\cal S}U^{\dagger} = {\cal S'}$.} there is a $Q
\in C^l$ and an invertible $R \in \Bbb{Z}_2^{n\times n}$ such that
\begin{equation}\label{QSR}QSR=S'.\end{equation} Note that the physical operation which transforms $|\psi\rangle$
into $|\psi'\rangle$ is entirely determined by $Q$; the right matrix multiplication with $R$ is
just a basis change within the stabilizer of the target state. 

\section{Reduction to graph states}

In this section we show that, under the transformations $S \to QSR$, each stabilizer generator
matrix $S$ can be brought into a (nonunique) standard form which corresponds to the generator
matrix of a graph state.

\noindent\textbf{Theorem 1}: {\it Each stabilizer state is equivalent to a graph state under local
Clifford operations.}

\noindent{\it Proof}: Consider an arbitrary stabilizer with generator matrix $S = \left [ \begin{array}{c} Z\\
X \end{array}\right].$ The result is obtained by proving the existence of a local Clifford
operation $Q \in C^l$ such that $QS=\left [ \begin{array}{c} Z'\\  X'
\end{array}\right]$ has an invertible lower block $X'$. Then \[S' := QSX'^{-1} = \left [ \begin{array}{c} Z'X'^{-1}\\
I \end{array}\right],\] where $Z'X'^{-1}$ is symmetric from the property $S'^TPS'=0$; furthermore,
the diagonal entries of $Z'X'^{-1}$ can be put to zero by additionally applying  the operation
$\left [
\begin{array}{cc} 1&1\\ 0&1
\end{array}\right]$ to the appropriate qubits, since this
operation flips the $i$th diagonal entry of $Z'X'^{-1}$ when applied on the $i$th qubit. Eventually
we end up with a graph state generator matrix of the desired standard form.

We now construct a local Clifford operation $Q$ that yields an invertible lower block $X'$. We
start by performing a basis change in the original stabilizer in order to bring $S$ in the form
\[S \to \left [ \begin{array}{cc} R_z&S_z\\R_x &0
\end{array}\right],\]
such that $R_x$ is a full rank $n\times k$-matrix, where $k=\mbox{ rank }X$; the blocks $R_z$,
$S_z$ have dimensions $n\times k$, resp. $n\times(n-k)$. The symplectic self-orthogonality of the
stabilizer implies that $S_z^T R_x = 0$. Furthermore, since $S_z$ has full rank, it follows that
the column space of $S_z$ and the column space of $R_x$ are each other's orthogonal complement.

Now, as $R_x$ has rank $k$, it has an invertible $k \times k$-submatrix. Without loss of
generality, we assume that the matrix consisting of the first $k$ rows of $R_x$ is invertible, i.e.
$R_x = \left[ \begin{array}{c} R_x^1\\R_x^2
\end{array}\right],$ where the upper $k\times k$-block $R_x^1$ is invertible  and $R_x^2$ has
dimensions $(n-k)\times k$. Partitioning $S_z$ similarly in a $k\times (n-k)$-block $S_z^1$ and a
$(n-k)\times (n-k)$-block $S_z^2$ , i.e. $S_z = \left[ \begin{array}{c} S_z^1\\S_z^2
\end{array}\right],$ the property $S_z^T R_x  = 0$ then implies that $S_z^2$ is also invertible: for,
suppose that there exist $x \in \Bbb{Z}_2^{n-k}$ such that $(S_z^2)^T x=0$; then the
$n$-dimensional vector $v :=(0, \dots, 0, x)$ satisfies $S_z^T v =0$ and therefore $v= R_x y$ for
some $y \in \Bbb{Z}_2^{k}$. This last equation reads \[\left[
\begin{array}{c} 0\\x
\end{array}\right] = \left[ \begin{array}{c} R_x^1\\R_x^2
\end{array}\right]y = \left[ \begin{array}{c} R_x^1y\\R_x^2y
\end{array}\right].\] Since $R_x^1$ is by construction invertible, $R_x^1y =0$ implies that $y=0$, yielding $x =R_x^2y
=0$. This proves the invertibility of $S_z^2$.

In a final step, we perform a Hadamard transformation $ \left [
\begin{array}{cc} 0&1\\ 1&0 \end{array}\right]$ on the qubits $k+1, \dots, n$. It is now easy to
verify that this operation indeed yields an invertible lower $n\times n$-block in the new generator
matrix, thereby proving the result. \hfill $\square$

This proposition is a special case of a result by Schlingemann \cite{stabgraphcode}, who showed, in
a more general context of $d$-level systems rather then qubits, that each stabilizer code is
equivalent to a graph code.

\noindent\textbf{Remark}: {\it overall phases - } Theorem 1 implies that our disregard of the
overall phases of the stabilizer elements is justified. Indeed, this result states that each
stabilizer state is equivalent to some graph state $|\psi_{\mu_1\mu_2\dots\mu_n}(G)\rangle$, for
some $\mu_i$. As such a state is equivalent to the state $|G\rangle$, there is no need to keep
track of the phases.

Theorem 1 shows that we can restrict our attention to graph states when studying the local
equivalence of stabilizer states. Note that in general the image of a graph state under a local
Clifford operation $Q = \left [ \begin{array}{cc} A&B\\
C&D \end{array}\right]$ need not again yield another graph state, as this transformation maps
\begin{equation}\label{image}\left [ \begin{array}{c} \theta\\
I \end{array}\right] \to Q \left [ \begin{array}{c} \theta\\
I \end{array}\right] = \left [ \begin{array}{c} A\theta+B\\
C\theta+D \end{array}\right]\end{equation} for $\theta \in \Theta$. The image in (\ref{image}) is
the generator matrix of a graph state if and only if $(a)$ the matrix $ C\theta +D$ is nonsingular
and $(b)$ the matrix $\theta':= \left( A\theta + B\right) \left(
C\theta +D \right)^{-1}$ has zero diagonal. Then \[Q \left [ \begin{array}{c} \theta\\
I \end{array}\right] \left( C\theta +D \right)^{-1} = \left [ \begin{array}{c} \theta'\\
I \end{array}\right]\] is the generator matrix for a graph state with adjacency matrix
$\theta'\in\Theta$. Note that we need not impose the constraint that $\theta'$ be symmetric, since
this is automatically the case,
as $\left [ \begin{array}{c} \theta'\\
I \end{array}\right]$ is the image of a stabilizer generator matrix under a Clifford operation, and
thus \[\left [ \begin{array}{cc} {\theta'}^T& I \end{array}\right] P \left [ \begin{array}{c} \theta'\\
I \end{array}\right] =0.\] These considerations lead us to introduce, for each $Q \in C^l$, a
domain of definition $\mbox{dom}(Q)$, which is the set consisting of all $\theta \in \Theta$ which
satisfy the  conditions $(a)$ and $(b)$. Seen as a transformation of the space $\Theta$ of all
graph state adjacency matrices, $Q$ then maps $\theta \in \mbox{dom}(Q)$ to
\begin{equation}\label{Q(theta)} Q(\theta):= \left( A\theta + B\right) \left( C\theta +D
\right)^{-1}.
\end{equation}
In this setting, it is of course a natural question to ask how the operations (\ref{Q(theta)})
affect the topology of the graph associated with $\theta$. We tackle this problem in the next
section.

To conclude this section we state and prove a lemma which we will need later on in the paper.

\noindent\textbf{Lemma 1}: {\it Let $\theta \in \Theta$ and $C, D$ be diagonal matrices s.t.
$C\theta + D$ is
invertible. Then there exists a unique $Q := \left [ \begin{array}{cc} A&B\\
C&D \end{array}\right] \in C^l$, where $A, B$ are diagonal matrices, such that $\theta \in \mbox{
dom}(Q)$.}

\noindent{\it Proof}: Note that, since $C\theta + D$ is invertible, we only need to look for a $Q$
s.t. $Q(\theta)$ has zero diagonal in order for $\theta$ to be in the domain of $Q$. First we will
prove the uniqueness of $A$ and $B$: suppose there exist two pairs of diagonal matrices $A, B$ and
$A', B'$
s.t. \[Q := \left [ \begin{array}{cc} A&B\\
C&D \end{array}\right], Q':= \left [ \begin{array}{cc} A'&B'\\
C&D \end{array}\right] \in C^l \] and $\theta \in \mbox{ dom}(Q)$, $\theta \in \mbox{ dom}(Q')$.
Denoting $\theta_z:= A\theta+B$, $\theta_z':= A'\theta+B'$ and $\theta_x:= C\theta+D$, we have
$Q(\theta) = \theta_z\theta_x^{-1}$ and $Q'(\theta) = \theta_z'\theta_x^{-1}$. Now, denoting by
$z_i^T, {\bar z_i}^T, x_i^T$ the rows of resp. $\theta_z, \theta_z', \theta_x$, the crucial
observation is that either ${\bar z_i}^T= z_i^T$ or ${\bar z_i}^T= z_i^T + x_i^T$ for all $i = 1,
\dots, n$, which is a direct consequence of the fact that $Q, Q'$ have the same lower blocks $C,
D$. Now, if the latter of the two possibilities is the case for some $i_0$, the $i_0$th diagonal
entries of $Q(\theta)$ and $Q'(\theta)$ must be different, since $Q'(\theta)_{i_0i_0} = \bar
z_{i_0}^T (\theta_x^{-1})_{i_0} = z_{i_0}^T (\theta_x^{-1})_{i_0} + x_{i_0}^T(\theta_x^{-1})_{i_0}
= Q(\theta)_{i_0i_0} + 1$, with $(\theta_x^{-1})_{i_0}$ the $i_0$th column of $\theta_x^{-1}$. As
both $Q(\theta)$ and $Q'(\theta)$ have zero diagonal, this yields a contradiction and we have
proven the uniqueness of $A$ and $B$. To prove existence, note that for every $i$, there are
exactly two couples
$(a_{i}, b_{i})$ s.t. $\left[ \begin{array}{cc} a_{i}&b_{i}\\
C_{ii}&D_{ii} \end{array}\right]$ is invertible. It follows from the above argument that we can
always tune $(a_{i}, b_{i})$ such that $(A\theta+B)(C\theta+D)^{-1}$ has zero diagonal, where we
take $A_{ii}=a_i$ and $B_{ii}=b_i$ for $i=1, \dots, n$. Since each $2\times2$-matrix
$\left [ \begin{array}{cc} A_{ii}&B_{ii}\\
C_{ii}&D_{ii} \end{array}\right]$ is invertible, the matrix $Q = \left [ \begin{array}{cc} A&B\\
C&D \end{array}\right]$ is an element of $C^l$, which proves the result. \hfill $\square$

\section{Local Clifford operations as graph transformations}

In this section, we investigate how the transformations (\ref{Q(theta)}) can be translated as graph
transformations. First we need some graph theoretical notions: two vertices $i$ and $j$ of a graph
$G = (V, E)$ are called adjacent vertices, or neighbors, if $\{i,j\} \in E$. The neighborhood $N(i)
\subseteq V$ of a vertex $i$ is the set of all neigbors of $i$. A graph $G' = (V', E')$ which
satisfies $V'\subseteq V$ and $E'\subseteq E$ is a subgraph of $G$ and one writes $G' \subseteq G$.
For a subset $A \subseteq V$ of vertices, the induced subgraph $G[A] \subseteq G$ is the graph with
vertex set $A$ and edge set $\{\{i, j\} \in E| i, j \in A\}$. If $G$ has an adjacency matrix
$\theta$, its complement $G^c$ is the graph with adjacency matrix $\theta + \Bbb{I}$, where
$\Bbb{I}$ is the $n\times n$-matrix which has all ones, except for the diagonal entries which are
zero.

\noindent\textbf{Definition 1}: For each $i=1, \dots, n$, the graph transformation $g_i$ sends an
$n$-vertex graph $G$ to the graph $g_i(G)$, which is obtained by replacing the subgraph $G[N(i)]$,
i.e. the induced subgraph of the neigborhood of the $i$th vertex of $G$, by its complement. In
terms of adjacency matrices, $g_i$ maps $\theta \in \Theta$ to \[g_i(\theta) = \theta +
\theta\Lambda_i\theta + \Lambda, \] where $\Lambda_i$ has a 1 on the $i$th diagonal entry and zeros
elsewhere and $\Lambda$ is a diagonal matrix such as to yield zeros on the diagonal of
$g_i(\theta)$.

The transformations $g_i$ are obviously their own inverses. Note that in general different $g_i$
and $g_j$ do not commute; however, if $\theta \in \Theta$ has $\theta_{ij}=0$, it holds that
$g_ig_j(\theta)=g_jg_i(\theta)$, as one can easily verify.

\noindent\textit{Example}: Consider the $5$-vertex graph $G$ whith adjacency matrix $\theta_{ij}=1$
for all $i\neq j$ and $\theta_{ii}=0$ for all $i$ (i.e. the complete graph), which is the defining
graph for the GHZ state. The application of the elementary local Clifford operation $g_1$ to this
graph is shown in Fig. 1.

\begin{figure}
\includegraphics{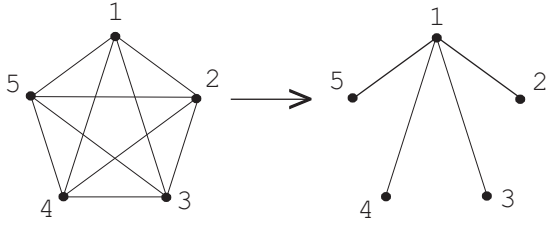}
\caption{\label{fig:epsart} Application of the graph operation $g_1$ to the GHZ graph.}
\end{figure}

The operations $g_i$ can indeed be realized as local Clifford operations (\ref{Q(theta)}). This is
stated in theorem 2 and was found independently by Hein {\it et al.} \cite{entgraphstate}.


\noindent\textbf{Theorem 2}: {\it Let $g_i$ be defined as before and $\theta \in \Theta$. Then
\[g_i(\theta) = Q_i (\theta),\] where \[ Q_i = \left [
\begin{array}{cc} I& \mbox{diag}(\theta_i)\\ \Lambda_i& I \end{array}\right] \in C^l,\] where
$\mbox{diag}(\theta_i)$ is the diagonal matrix which has $\theta_{ij}$ on the $j$th diagonal entry,
for $j = 1, \dots ,n$.}

\noindent{\it Proof}: The result can be shown straightforwardly by calculating $Q_i (\theta) =
(\theta + \mbox{ diag}(\theta_i))(\Lambda_i \theta + I)^{-1}$ and noting that the matrix $\Lambda_i
\theta + I$ is its own inverse for any $\theta$. \hfill $\square$


The remainder of this section is dedicated to proving that the operations $g_i$ in fact generate
the entire orbit of a graph state under local Clifford operations, that is to say, two graph states
$|G\rangle, |G'\rangle$ are equivalent under the local Clifford group iff there exists a finite
sequence $g_{i_1}, \dots, g_{i_N}$ such that $g_{i_N} \dots g_{i_1}(G)=G'$. This result completely
translates the action of local Clifford operations on graph states into a corresponding action on
their graphs. In order to
prove the result, we need the following lemma.

\noindent\textbf{Lemma 2}: {\it Define the matrix class $\Bbb{T} \subseteq \Bbb{Z}_2^{n\times n}$
by
\begin{eqnarray}\Bbb{T} = &&\left\{ C\theta + D| \ \theta \in \Theta \mbox{ and } C, D \mbox{ are diagonal}\right.
\nonumber\\ && \left. \mbox{ and } C\theta + D \mbox{ is invertible}
\right\}\nonumber\end{eqnarray} and consider an  element $R \in \Bbb{T}$. Choose $\theta, C, D$
such that $R=C\theta + D$. Define the transformation $f_i$ of $\Bbb{T}$ by $f_i(X) = X(\Lambda_i X
+ X_{ii}\Lambda_i + I)$, for $X\in\Bbb{T}$, and denote $f_{jk}(\cdot):= f_jf_kf_j(\cdot)$. Then (i)
there exists a finite sequence of $f_{i}$'s and $f_{jk}$'s such that
\begin{equation}\label{sequence} f_{j_Mk_M}\dots f_{j_1k_1}f_{i_N}\dots f_{i_1} (R) =I,
\end{equation} where all the indices in the sequence
are different;  (ii) there exists a unique $Q_0 = \left [ \begin{array}{cc} A_0&B_0\\
C&D
\end{array}\right] \in C^l$, such that \ $\theta \in \mbox{ dom}(Q_0)$ and \[ g_{j_Mk_M}\dots
g_{j_1k_1}g_{i_N}\dots g_{i_1} (\theta) =Q_0(\theta), \] where $g_{jk}(\cdot):= g_jg_kg_j(\cdot)$.}

\noindent\textit{Proof}: First, straightforward calculation shows that $f_i$ maps the class of
matrices of the form $C\theta+D$ to itself. Furthermore, for each $X\in\Bbb{T}$ the matrix
$\Lambda_i X +X_{ii}\Lambda_i + I$ is invertible, which implies that $f_i$ maps invertible matrices
to invertible matrices. Therefore each $f_i$ is indeed a transformation of $\Bbb{T}$. Now,
statement $(i)$ is proven by applying the algorithm below, where the idea is to successively make
each $i$th row of $R$ equal to the $i$th cannonical basis vector $e_i^T= [0\dots 0\ 1\ 0 \dots 0]$,
by applying the correct $f_j$'s in each step. The image of $R$ throughout the consecutive steps
will be denoted by the same symbol $\bar R = \left(r_{ij}\right)$. Now, the algorithm consists of
repeatedly performing one of the two following sequences of operations on $\bar R$:

\textit{Case 1}: If $\bar R$ has a diagonal entry $r_{i_0i_0}=1$ (and the $i_0$th row of $\bar R$
is not yet equal to the basis vector $e_{i_0}^T$), apply $f_{i_0}$. It is easy to verify that, in
this situation, $f_{i_0}$ transforms the $i_0$th row of $\bar R$ into the basis vector $e_{i_0}^T$.

\textit{Case 2}: If the conditions for case 1 are not fulfilled, apply the following sequence of
three operations: firstly, fix a $j_0$ such that $r_{j_0j_0}=0$ and apply $f_{j_0}$. It can easily
be seen that then diag$(\bar R)$ $\to$ diag$(\bar R) + {\bar R} _{j_0}$, where ${\bar R} _{j_0}$ is
the $j_0$th column of $\bar R$ and diag$(\bar R)$ is the diagonal of $\bar R$. Since $\bar R$ is
invertible, ${\bar R} _{j_0}$ has some nonzero element, say on the $k_0$th position $r_{k_0j_0}=1$.
Therefore, the application of $f_{j_0}$ has put a $1$ on the $k_0$th diagonal entry of the
resulting $\bar R$. Now we apply $f_{k_0}$, turning the $k_0$th row into $e_{k_0}^T$ as in case 1.
Furthermore, this second operation has put $r_{j_0k_0}$ on the $j_0$th diagonal entry and, from the
symmetry of $\bar R$, it holds that $r_{j_0k_0}=r_{k_0j_0}=1$. Therefore, by again applying
$f_{j_0}$, we obtain an $e_{j_0}^T$ on the $j_0$th row of the resulting $\bar R$. Finally, we note
that after performing this sequence of operations, we end up with an $\bar R$ which will again
satisfy the conditions for case 2.

Repetition of these elementary steps will eventually yield the identity matrix, which concludes the
proof of statement $(i)$.

Statement $(ii)$ is proven by induction on the length of the sequence of $f_i$'s and $f_{jk}$'s. As
it turns out,  the easiest way to do this is to consider the $f_i$'s and $f_{jk}$'s as two
different types of elementary blocks in the sequence (\ref{sequence}). The proof will therefore
consist of two parts {\it A} and {\it B}, part {\it A} dealing with the $f_i$'s and part {\it B}
with the $f_{jk}$'s.

\underline{Part {\it A}:} in the basis step of the induction we have $f_i(R) = I$, where
$R\in\Bbb{T}$.  Any such $R$ satisfies $R = \Lambda_i R + R_{ii}\Lambda_i + I$ and therefore must
be of the form
\[R = \left[
\begin{array}{ccccccc} 1& & & & & &  \\& \ddots & & & & &  \\& & 1& & & &  \\x_{1}& & x_{i-1}
&1&x_{i+1} &  &x_{n}  \\& & & & 1& &
\\& & & &  & \ddots&   \\& & & &&  &1   \end{array}\right] \quad \leftarrow i\]
for some $x_{j}$. Then any $\theta, C, D$ which satisfy $R = C\theta + D$ must satisfy $\theta_{ij}
= x_j$ and $D = I$; moreover, if $C_{jj} =1$ for $j \neq i$ then the $j$th row of $\theta $ must be
equal to zero and if $C_{ii} =0$ the $i$th row of $\theta $ must be to zero. It is now easy to see
that $g_i(\theta) = Q(\theta)$, with $Q = \left [ \begin{array}{cc} \cdot &\cdot\\
C&D
\end{array}\right]$.

In the induction step, we suppose that the statement holds for all sequences $f_{i_1}, \dots,
f_{i_N}$ of fixed length $N$ and prove that this implies that the statement is true for sequences
of length $N+1$. We start from the given that \[f_{i_N} \dots f_{i_1} f_i(R) = I\] for some $f_i,
f_{i_1}, \dots, f_{i_N}$ and $R\in\Bbb{T}$ and we choose $\theta, C, D$ such that $C\theta+D = R$.
Note that it follows from case 1 in the algorithm in part (i) of the lemma that we may take $C_{ii}
= 1 = D_{ii}$, as a single $f_i$ (as opposed to an $f_{ij}$) is only applied when $R_{ii} = 1$.
Furthermore, we will denote by $\omega$ the set of all $j \in \{1,\dots,n\}$ such that $C_{jj}=1$.
As $R$ is invertible, this implies that $D_{kk}=1$ for $k\in \omega^c$. Now, denoting $R':=
f_i(R)$, we have $f_{i_N} \dots f_{i_1}(R') = I$, which allows us to use the result for length $N$:
for any $\theta', C', D'$ that satisfy $C'\theta'+D' = R'$ there exists a $Q'\in C^l$ which has
$C', D'$ as its lower blocks such that $\theta' \in \mbox{ dom}(Q')$ and
\begin{equation}\label{stap}g_{i_N}\dots g_{i_1} (\theta') =Q'(\theta').\end{equation} We make the
following choices for $\theta', C', D'$:
\begin{eqnarray}
\theta'&=&g_i(\theta)\nonumber\\
C' &=& C + \Lambda_i\nonumber\\
D'&=&D + \mbox{diag}(\theta_i)_{\omega}\nonumber
\end{eqnarray}
where $\mbox{diag}(\theta_i)_{\omega}$ is the diagonal matrix which has $\theta_{ij}$ on the $j$th
diagonal entry if $j\in\omega$ and zeros elsewhere. This choice for $\theta', C', D'$ indeed yields
$C'\theta'+D' = R'$; we will however omit the calculation since it is straightforward. Now, using
the definition of $\theta'$ and Theorem 2, equation (\ref{stap}) becomes
\begin{equation}g_{i_N}\dots g_{i_1} g_i(\theta) =(Q'Q_i)(\theta).\end{equation} It is now again
straightforward to show that $Q:= Q'Q_i$ has $C$ and $D$ as its lower blocks. Uniqueness follows
from lemma 1. This proves the induction step, thereby concluding the proof of part $A$.

\underline{Part {\it B}:} The proof of this part is analogous to part $A$, though a bit more
involved. The basis step now reads $f_{jk}(R)=I$.  Now case 2 in the algorithm in part (i) of the
lemma implies that $R_{jj}=0=R_{kk}$ and $R_{jk}=1$, as only if this is the case, $f_{jk}$ is
applied in the algorithm. For simplicity, but without losing generality, we take $i=1, j=2$. Then
$R$ must be of the form \[R =\left[\begin{array}{cc|c}
0 & 1& \quad \theta_1^T\quad \\
1& 0 & \quad \theta_2^T\quad \\
\hline & &\\
0 & 0& I\\
& &
\end{array}\right],\] where the $\theta_i$ are $(n-2)$-dimensional column vectors. Choosing $\theta, C, D$
s.t. $R=C\theta+D$, the matrix $\theta$ must satisfy \[\theta = \left[\begin{array}{cc|c}
0 & 1& \quad \theta_1^T\quad \\
1& 0 & \quad \theta_2^T\quad \\
\hline & &\\
\theta_1 & \theta_2& \phi\\
& &
\end{array}\right],\] where $\phi$ is a symmetric $(n-2)\times(n-2)$-matrix with zero diagonal;
furthermore $D_{11}=0=D_{22}$, $D_{jj} = 1$, $C_{11} = 1 = C_{22}$ and if $C_{j+2,j+2}=1$ then
$\theta_{1j} = \theta_{2j} = \phi_{kj}=0$, for $j,k=1, \dots, n-2$. We will give the proof for
$C_{j+2,j+2}=0$, the other cases are similar. Thus, we have to show that there exists a $Q_0\in
C^l$ with lower blocks $C, D$ s.t. $g_{12}(\theta) = Q_0(\theta)$. To prove this, we use theorem 2,
yielding \[Q_0 = \left [ \begin{array}{cc} \cdot &\cdot\\
\Lambda_1& I
\end{array}\right]\left [ \begin{array}{cc} I &\mbox{diag}(g_1(\theta)_2)\\
\Lambda_2& I
\end{array}\right]\left [ \begin{array}{cc} I & \mbox{diag}(\theta_1)\\
\Lambda_1& I
\end{array}\right],\] where $g_1(\theta)_2 =(1, 0, \theta_{13} + \theta_{23}, \dots, \theta_{1n} + \theta_{2n})$
 is the second column of $g_1(\theta)$. A simple calculation reveals that
 \[Q_0 = \left [ \begin{array}{cc} \cdot &\cdot\\
\ \Lambda_1+\Lambda_2& I+\Lambda_1+\Lambda_2
\end{array}\right] = \left[\begin{array}{cc} \cdot &\cdot\\
C& D
\end{array}\right],\] which proves the basis of the induction.

In the induction step, we again follow an analogous reasoning to part $A$: we suppose that the
statement is true for sequences $f_{j_1k_1}, \dots, f_{j_Nk_N}$ of length $N$ and prove that this
implies the statement for length $N+1$. Our starting point is now \[f_{j_Nk_N} \dots
f_{j_1k_1}f_{jk}(R)=I\] for some $f_{jk}, f_{j_1k_1}, \dots, f_{j_Nk_N}$ and $R\in \Bbb{T}$. Note
that again we have $R_{jj}=0=R_{kk}$ and $R_{jk}=1$ as in the basis step.  As from this point on
the strategy is identical as in part $A$ and all calculations are straightforward, we will only
give a sketch: first we denote $R' = f_{jk}(R)$; for $\theta, C, D$ s.t. $R=C\theta+ D$, we define
\begin{eqnarray}
\theta'&=&g_{jk}(\theta)\nonumber\\
C' &=& C + \Lambda_j\nonumber + \Lambda_k\\
D'&=&D + \Lambda_j\nonumber + \Lambda_k
\end{eqnarray}
It then straightforward to show that $R'=C'\theta'+D'$. The induction yields a $Q'\in C^l$ with
lower blocks $C', D'$ such that \[g_{j_Nk_N} \dots g_{j_1k_1}(\theta')=Q'(\theta').\] Using theorem
2, we calculate $Q_{jk}$ s.t. $g_{jk}(\theta) = Q_{jk}(\theta)$. Then \[g_{j_Nk_N} \dots
g_{j_1k_1}g_{jk}(\theta)=(Q'Q_{jk})(\theta)\] and a last calculation shows that $Q_0:=Q'Q_{jk}$ has
lower blocks $C$ and $D$. Uniqueness again follows from lemma 1. This proves part B of the lemma.
\hfill $\square$

The main result of this paper is now an immediate corollary of lemmas 1 and 2:

\noindent \textbf{Theorem 3}: {\it Let $\theta \in \Theta$. Then the operations $g_1, \dots, g_n$
generate the orbit of  $\theta$ under the action (\ref{Q(theta)}) of the local Clifford group
$C^l$.}

\noindent\textit{Proof}: Let $Q = \left [ \begin{array}{cc} A&B\\
C&D \end{array}\right] \in C^l$ such that $\theta \in \mbox{ dom}(Q)$. Now, as $C\theta + D$ is an
invertible element of $\Bbb{T}$, lemma 2(ii) can be applied, yielding a unique $Q_0 =
\left [ \begin{array}{cc} A_0&B_0\\
C&D \end{array}\right]\in C^l$ and sequence of $g_i$'s and $g_{jk}$'s such that $\theta \in \mbox{
dom}(Q_0)$ and \[ g_{j_Mk_M}\dots g_{j_1k_1}g_{i_N}\dots g_{i_1} (\theta) =  Q_0 (\theta),\] As $Q$
and $Q_0$ have the same lower blocks $C$ and $D$ and $\theta$ is in both of their domains, it
follows from lemma 1 that $Q_0 = Q$ and the result follows. \hfill $\square$

\section{Discussion}

The result in Theorem 3 of course facilitates generating the equivalence class of a given graph
state under local Clifford operations, as one only needs to successively apply the rule to an
initial graph. Note that the lemma 2 implies that one only needs to consider sequences
$g_{j_Mk_M}\dots g_{j_1k_1}g_{i_N}\dots g_{i_1}$ of limited length. Furthermore, the translation of
the operations (\ref{Q(theta)}) into sequences of elementary graph operations gets rid of annoying
technical domain questions.  It is important to notice that we have \emph{not} proven that each $Q
\in C^l$ corresponds directly to a sequence of $g_i$'s, since, in theorem 3, the decomposition into
$g_i$'s depends both on $Q$ as well as $\theta$.

In a final note, we wish to point out that testing whether two stabilizer states with generator
matrices $S, S'$ are equivalent under the local Clifford group is an easily implementable algorithm
when one uses the binary framework. Indeed, one has equivalence iff there exists a $Q \in C^l$ s.t.
\begin{equation}\label{algo}S^TQ^TPS' =0,\end{equation} as this expression states that the stabilizer subspaces generated by the
matrices $S'$ and $QS$ are orthogonal to each other with respect to the symplectic inner product.
Since any stabilizer subspace is its own symplectic orthogonal complement, the spaces generated by
$S'$ and $QS$ must be equal, which implies the existence of an invertible $R$ s.t. $S' = QSR$.
Equation
(\ref{algo}) is a system of $n^2$ linear equations in the $4n$ entries of $Q=\left [ \begin{array}{cc} A&B\\
C&D \end{array}\right]$, with $n$ additional quadratic constraints $A_{ii}D_{ii} + B_{ii}C_{ii} =1$
which state that $Q \in C^l$; these equations can be solved numerically by first solving the linear
equations and disregarding the constraints and then searching the solution space for a $Q$ which
satisfies the constraints. Although we cannot exclude that the worst case number of operations is
exponential in the number of qubits, in the majority of cases this algorithm gives a quick
response, as for large $n$ the system of equations is highly overdetermined and therefore
generically has a small space of solutions. Note that, when equivalence occurs, the algorithm
provides an explicit $Q$ wich performs the transformation.

\section{Conclusion}

In this paper, we have translated the action of local unitary operations within the Clifford group
on graph states into transformations of their associated graphs. We have shown that there is
essentially one elementary graph transformation rule, successive application of which generates the
orbit of any graph state under the action of local Clifford operations. This result is a first step
towards characterizing the LU-equivalence classes of stabilizer states.

\begin{acknowledgments}
M. Van den Nest thanks H. Briegel for inviting him to the Ludwig-Maximilians-Universit\"at in
Munich for collaboration and acknowledges interesting discussions with M. Hein and H. Briegel. This
research is supported by: Research Council KUL: GOA-Mefisto 666, several PhD/postdoc \& fellow
grants; Flemish Government: FWO: PhD/postdoc grants, projects, G.0240.99 (multilinear algebra),
G.0407.02 (support vector machines), G.0197.02 (power islands), G.0141.03 (Identification and
cryptography), G.0491.03 (control for intensive care glycemia), G.0120.03 (QIT), research
communities (ICCoS, ANMMM); AWI: Bil. Int. Collaboration Hungary/Poland; IWT: PhD Grants, Soft4s
(softsensors); Belgian Federal Government: DWTC (IUAP IV-02 (1996-2001) and IUAP V-22 (2002-2006),
PODO-II (CP/40: TMS and Sustainability); EU: CAGE; ERNSI; Eureka 2063-IMPACT; Eureka 2419-FliTE;
Contract Research/agreements: Data4s, Electrabel, Elia, LMS, IPCOS, VIB; M. Van den Nest
acknowlegdes support by the European Science Foundation (ESF), Quprodis and Quiprocone.
\end{acknowledgments}

\bibliography{localequivgraph}

\end{document}